Piotr KRASNOWSKI*, Jerome LEBRUN** and Bruno MARTIN**

# EXCHANGING KEYS WITH AUTHENTICATION AND IDENTITY PROTECTION FOR SECURE VOICE COMMUNICATION WITHOUT SIDE-CHANNEL

**ABSTRACT:**

Motivated by an increasing need for privacy-preserving voice communications, we investigate here the original idea of sending encrypted data and speech in the form of pseudo-speech signals in the audio domain. Being less constrained than military "Crypto Phones" and allowing genuine public evaluation, this approach is quite promising for public unsecured voice communication infrastructures, such as 3G cellular network and VoIP.

A cornerstone of secure voice communications is the authenticated exchange of cryptographic keys with sole resource the voice channel, and neither Public Key Infrastructure (PKI) nor Certificate Authority (CA). In this paper, we detail our new robust double authentication mechanism based on signatures and Short Authentication Strings (SAS) ensuring strong authentication between the users while mitigating errors caused by unreliable voice channels and also identity protection against passive eavesdroppers. As symbolic model, our protocol has been formally proof-checked for security and fully validated by Tamarin Prover.

## 1. INTRODUCTION

A growing privacy concern in voice communications triggers the development of secure VoIP communicators. Classical in-phone solutions are insecure against the potential spying malware installed on smartphones and military-grade Crypto-Phones remain a "niche" market suffering from high costs and constrained usability.

Figure 1 depicts an alternative scheme for secure speech communication, fully detailed, implemented and tested in [3]. Firstly, the original speech acquired at the headset's microphone is processed and enciphered by a voice encrypting (VE) device to produce in real-time a corresponding pseudo-speech signal in the audio domain. This secure encrypted pseudo-speech signal is transmitted as analog audio input to a phone for transmission through 3G/4G or VoIP.

---

* p.g.krasnowski@gmail.com (grants DGA/DS/MRIS nº 01D17022178 and AID nº SED0456JE75)
** Université Côte d'Azur and I3S-CNRS, Sophia Antipolis, France. {lebrun,bruno.martin}@i3s.unice.fr

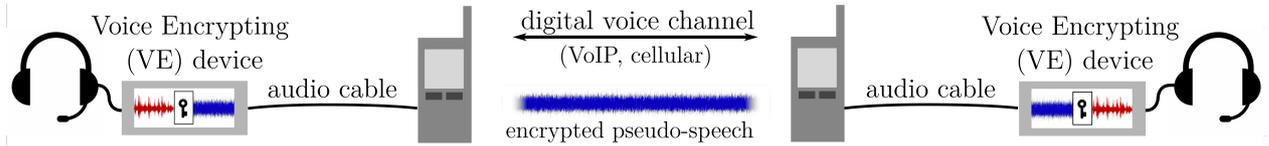

Fig. 1. Scheme of a secure voice communication over voice channels.

Likely distorted by the voice channel, the pseudo speech signal is received by the receiver phone. It is then deciphered by the external paired VE device on the other side of the channel. This proposed setting provides better protection against potential malware and greater usability, quite useful in unreliable environments lacking confidential communication infrastructure as for diplomatic and military services, journalists, lawyers, and traders.

To generate secure pseudo-speech signals that are robust to the typical distortions introduced by perceptually-oriented voice channels and to achieve real-time transmission of encrypted speech over vocal links are challenges tackled either through Data over Voice (DoV) [1,2] or using dedicated speech encryption algorithms [3].

The other major issue is the secure exchange of session keys between the VE devices, that can only be made through the same point-to-point voice channel, preventing the comfort of online CA to facilitate users' authentication. Moreover, the exchange must be robust against channel errors and message suppressions. Even well-established protocols such as ZRTP [4] prove to be overly complex and unsuitable.

In [5], we introduced a solution based on an Ephemeral Elliptic-Curve Diffie-Hellman Exchange (ECDHE) [5] that authenticates users using signatures and Short Authentication Strings (SAS), i.e. short sequences of digits or words compared vocally by the peers at the end of the key exchange. This protocol also provides perfect forward secrecy and robustness against possible desynchronization caused by the adversary or by channel errors. We improve here upon this protocol by introducing additional identity protections against passive eavesdroppers by enciphering fixed user identifiers using ephemeral keys. We get the same requirements on the number of messages to be exchanged between the parties and similar robustness against desynchronization. We also detail a verification as a symbolic model using the Tamarin Prover [6] with conclusive results on the level of security and authentication.

## 2. THE PROTOCOL

The key exchange protocol for secure voice communication is detailed in Fig. 2. Initially Alice and Bob establish a non-encrypted call using their preferred voice application and perform an automatic role negotiation to prevent mutual interference.

Next, they exchange values that are used to derive the Session Key and the SAS. The user identifiers, nonces, and signatures are protected by a symmetric cipher with three ad-hoc secret keys. The senders' signature verification key can be optionally provided to the recipients by using a system of virtual cards before the communication starts. If the recipient does not obtain the verification key, the signature is not processed.

After all cryptographic parameters are successfully exchanged, voice encryption can be started. At this moment, each participant can request a vocal challenge of SAS equality with the peer. SAS comparison is obligatory if any of the users was not able to verify the signature. The SAS is displayed on the VE device as a short string of digits or words to be vocally uttered by the users and re-typed on the paired device. It is assumed that the comparison process gets authenticated when the users simultaneously validate by ear the voice characteristics of the peer (timbre, tempo, etc.).

The verification of the protocol was done in a symbolic model using the Prover by assuming perfect cryptographic primitives that give the adversary no advantage. The adversary is granted a full control over the network and the capability to steal signature keys. Several protocol restrictions were relaxed, allowing users to run multiple protocol instantiations at the same time and to revoke the verification key of the peer. Vocal comparison is modelled as communicating over an authenticated (not secret) channel, which the adversary can intercept but not modify. The Tamarin code is available online at https://osf.io/gujfp.

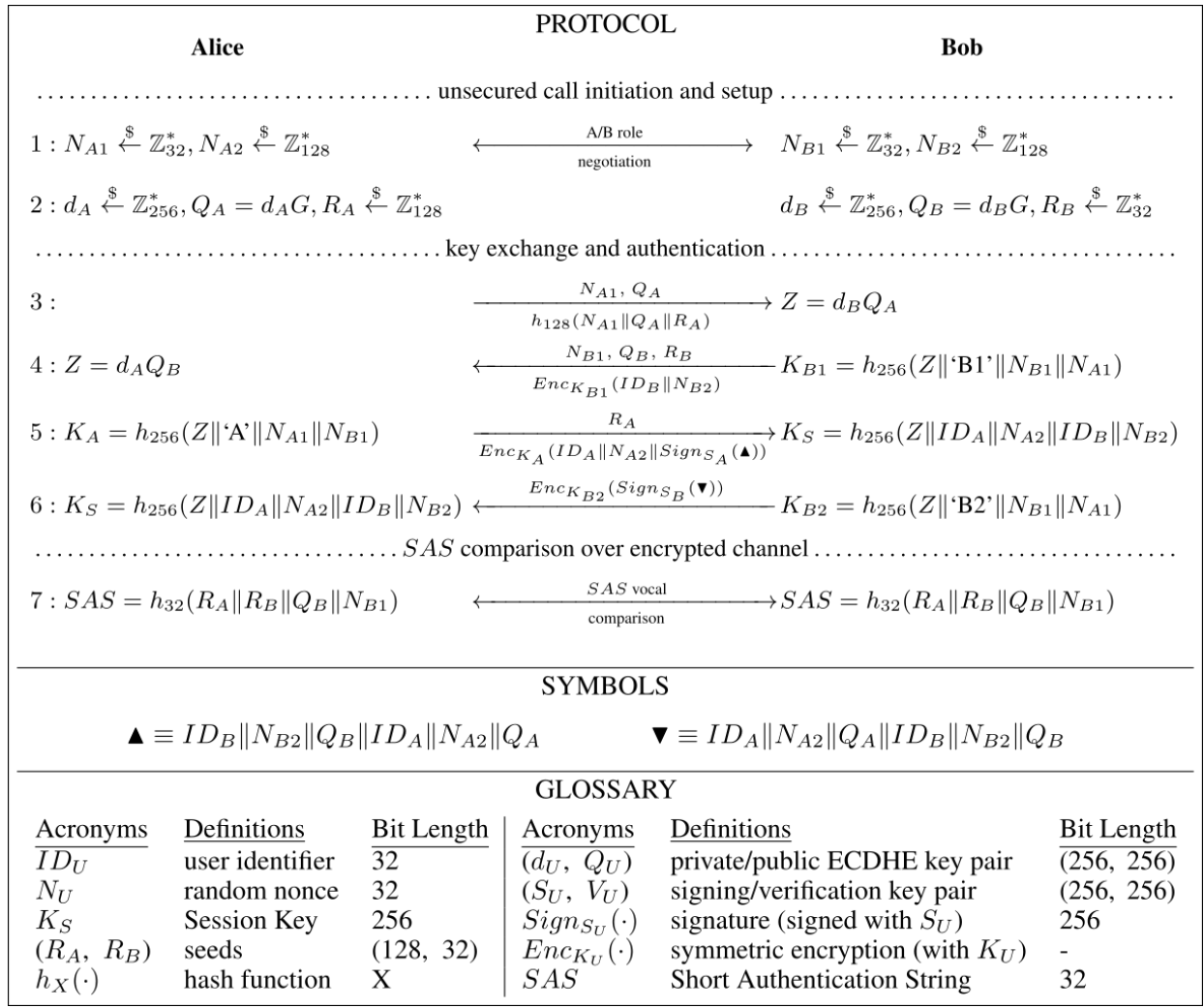

Fig. 2. Key exchange protocol with identity protection. The symbol '$' denotes a random choice.

Evaluation was done in four authentication configurations assuming the peers can univocally identify themselves by voice. The results in Table 1 indicate that either signatures or SAS comparison provide perfect forward secrecy and injective agreement, while lack of any authentication rescinds security. Thus, VE devices should strictly encourage SAS comparison whenever the signature verification is impossible.

Table 1. Security properties verified by Tamarin in four scenarios: (a) mutual signature authentication, (b) unilateral signature authentication, (c) SAS vocal verification and (d) no authentication.

| Authentication Scenario | (a) | (b) | (c) | (d) |
|---|---|---|---|---|
| Session Key secrecy | yes | yes | yes | no |
| Forward secrecy | yes | yes | yes | no |
| Injective agreement | yes | yes | yes | no |

## 3. CONCLUSION

We described and successfully verified a new key exchange protocol with authentication and identity protection for secure voice communication over voice channels. Future work is planned to implement our protocol on dedicated hardware and to test it over real-world voice channels for security and robustness against signal distortions.